\documentclass[sigconf,screen]{acmart}
\AtBeginDocument{%
  \providecommand\BibTeX{{%
    \normalfont B\kern-0.5em{\scshape i\kern-0.25em b}\kern-0.8em\TeX}}}

\usepackage{multirow}
\usepackage{tabularx}
\usepackage{subcaption}
\usepackage{makecell}

\begin{document}


% Change the title to reflect it is multi-institutional?
\title{Demystify, Use, Reflect: \\Preparing students to be informed LLM-users}


\author{Nikitha Donekal Chandrashekar}
\affiliation{%
  \institution{Virginia Tech}
  \city{Blacksburg}
  \country{USA}}
\email{nikitha@vt.edu}
 
\author{Sehrish Basir Nizamani}
\affiliation{%
  \institution{Virginia Tech}
  \city{Blacksburg}
  \country{USA}}
\email{sehrishbasir@vt.edu}

\author{Margaret Ellis}
\affiliation{%
  \institution{Virginia Tech}
  \city{Blacksburg}
  \country{USA}}
\email{maellis1@vt.edu}
 
\author{Naren Ramakrishnan}
\affiliation{%
  \institution{Virginia Tech}
  \city{Alexandria}
  \country{USA}}
\email{naren@vt.edu}

%%\author{Nikitha Donekal Chandrashekar, Sehrish Nizamani, Margaret Ellis, and Naren Ramakrishnan}
%%\affiliation{%
%%  \institution{Department of Computer Science}
%%  \city{Virginia Tech}
%%  \country{}}

\renewcommand{\shortauthors}{Anon et al.}

\begin{abstract}
We transitioned our post-CS1 course that introduces various subfields of computer science so that it integrates Large Language Models (LLMs) in a 
structured, critical, and practical manner.
It aims to help students develop the skills needed to engage meaningfully and responsibly with AI. 
The course now includes explicit instruction on how LLMs work, exposure to current tools, ethical issues, and activities that encourage student reflection on personal use of LLMs as well as the larger evolving landscape of AI-assisted programming.  
In class, we demonstrate the use and verification of LLM outputs, guide students in the use of LLMs as an ingredient in a larger problem-solving loop, and require students to disclose and acknowledge the nature and extent of LLM assistance.  
Throughout the course, we discuss risks and benefits of LLMs across CS subfields. 
In our first iteration of the course, we collected and analyzed data from students pre and post surveys. 
Student understanding of how LLMs work became more technical, and their verification and use of LLMs shifted to be more discerning and collaborative. 
These strategies can be used in other courses to prepare students for the AI-integrated future.
\end{abstract}

% %%
% %% The code below is generated by the tool at http://dl.acm.org/ccs.cfm.
% %% Please copy and paste the code instead of the example below.
% <concept_id>10003456.10003457.10003527.10003530</concept_id>
% <concept_desc>Social and professional topics~Model curricula</concept_desc>

% \ccsdesc[500]{Social and professional topics~Model curricula}

\keywords{Large Language Models, Generative AI, Problem Solving}

\maketitle
\section{Introduction}
To help students be more intentional about their use of LLMs, we updated our course to open up the `black box' and help them understand its inner workings.
In particular, we underscore how an LLM can be part of (but not replace) the iterative problem-solving and verification process. This process is critical, as these models function as a `leaky bucket'—capable of generating sophisticated code yet prone to basic errors. Our key contributions are:
\begin{enumerate}
    \item Emphasizing students' understanding of LLM internals and LLM pipelines in conjuntion with use of LLMs
    \item Designing assignments that require verification and multiple LLM iterations 
    \item Encouraging reflection on effective and ethical us of LLMs 
\end{enumerate}
Our approach demystifies LLMs so that CS students understand how they work and become aware of the risks and benefits of using LLMs across various CS subfields, including algorithm analysis and design, software engineering, networking, human-computer interaction, database systems, and machine learning. The goal is to build student capacity to use LLMs ethically and appropriately.

\section{Related Research}
Related research has focused on the use of LLMs by computer science students. As Vadaparty et al. ~\cite{vadaparty2024} found, students struggle to discern the appropriate use of LLMs even when they are allowed to use them for CS coursework.
This concern is echoed in CS1 research such as Margulieux et al. ~\cite{margulieux2024}, who report that while successful students can leverage LLMs to enhance their learning, struggling students may actually be hindered by their use. The study performed by Smith et al. \cite{smith2025} involves spotting AI missteps, where students critically analyze LLM-generated errors in CS1 courses, encouraging meta-cognitive skills and deeper engagement with debugging. Our course attempts to promote effective use of LLMs.

% Requires: \usepackage{multirow}
\begin{table*}[t]
    \centering
     \caption{Course Schedule for LLM Topics Added Across Weeks to Classwork (CW), Homework (HW), and Lectures (L).}
    \begin{tabular}{|l|c|c|c|c|c|c|c|c|c|c|c|c|c|}
      \hline
   & & \multicolumn{2}{c|}{\textit{LLMs}} & \multicolumn{2}{c|}{\textit{Algorithms}} & \multicolumn{2}{c|}{\textit{Soft. Eng.}} & \multicolumn{2}{c|}{\textit{Networking}} & \multicolumn{2}{c|}{\textit{HCI}} & \multicolumn{2}{c|}{\textit{Data Science}}   \\
    \hline
    
    {\textbf {Goals}} & {\textbf{Approaches}} & W1 & W2 & W3 & W4 & W5 & W6 & W8 & W9 & W10 & W11 & W12 & W13  \\
    \hline
    \multirow{2}{*}{Demystify LLMs} & \multicolumn{1}{|l|}{How LLMs Work} & CW & L &  &  & & &  L &  &  &  &  &  \\
    \cline{2-14}
    & \multicolumn{1}{|l|}{Current Resources} & HW  & CW & HW  &  &  HW &  &  & & & & & HW  \\
    \hline
    \multirow{2}{*}{LLM Usage Skills} & \multicolumn{1}{|l|}{Demos} & & & L & L & L & & L & & CW & L &  & \\
    \cline{2-14}
    & \multicolumn{1}{|l|}{Semi-LLM-Resistant} & &  &  & HW & HW & HW &  & HW &  &  &  & HW \\
    \hline
    \multirow{2}{*}{\makecell[l]{Reflective and\\ Responsible LLM Use}} & \multicolumn{1}{|l|}{Risks and Benefits} &  & L &  &    & CW, L &   L &  & L & &  L & &  \\
    \cline{2-14}
    & \multicolumn{1}{|l|}{Reflection on Use} & &  & L &    & HW, L & L &  &  &  &  & & HW \\
    \hline
    \end{tabular}
   
    \label{tab:llm_schedule}
\end{table*}

\section{Approach}

Our institution is a large public university in the United States. This is a required course for CS majors with CS1 as a prerequisite course. It introduces students to the technical aspects of LLMs and then to several subfields of computer science. For each of the subfields, students are provided with an introduction and the interplay of the subfield with LLMs, including risks and benefits. Students can use LLMs on the assignments and describe their usage. Students are asked to reflect on their experiences with LLMs to improve their metacognition. Our approaches overlap with those described in the literature—such as in-class demos, active learning strategies, LLM usage policies, and explicit instruction on LLMs~\cite{prather2025}.
Table~\ref{tab:llm_schedule} presents the new LLM-related content and assignments, organized week-by-week. Links to assignments, resources, sample code, and slides available at \textbf{anonymized url}. 

%\subsection{Demystifying LLMs}

To demystify LLMs, students are familiarized with them through adapted material from an upper-level machine learning course. The curriculum covers how, why, and when LLMs work. Throughout the semester, we have assignments on current topics related to LLMs such as prompting, guardrails, chatbots, computing ethics, bias, use of LLMs to code, and cheating on coding interviews.

%\subsection{LLM Usage Skills}
During lectures, instructors provide demonstrations of using LLM tools along with code and unit tests that are both human-written and LLM-written. Assignments, such as `Wannabe Palindrome', are intentionally structured so that students cannot rely on a single prompt or zero-shot solution; success depends on careful iteration and debugging. 

%\subsection{Reflective and Responsible LLM Use}

We delve into implications of using LLMs such as the risk that LLMs will be biased due to training data, and how LLMs can easily confuse complex business rules or real-world constraints. When teaching students the basics of the Internet, we discuss how data centers, the energy demands of AI, and the evolution of new models can present sustainability and privacy concerns. Throughout the semester, we recognize the importance of encouraging students to reflect on their experiences with LLMs ~\cite{tankelevitch2024}.

\section{Impact}

In the Spring of 2025 the course was taught by 3 instructors across 4 sections to 272 students who completed a pre and post survey designated as exempt by our Institutional Review Board. A total of 244 students consented for both the surveys.
We performed qualitative coding on the responses in both the surveys to the questions ``How do LLMs work?'', ``A LLM is \rule{0.3cm}{0.15mm} . '', ``How do you use LLMs effectively?'' and ``How do you assess the correctness of information from an LLM?'' to gauge the potential impact of our course updates. We followed an LLM-Assisted Thematic Analysis (TA) \cite{wang2025lata} approach, where we first performed preliminary coding on de-identified survey data using GPT-4. Then following Braun and Clarke's TA framework we derive the themes and patterns.

Our analysis reveals an increase in understanding with explanations of LLMs that involve data training, probabilistic outputs, and functional use. The usage of the term \textit{``token''} increased from 24 to 94 mentions. Students increasingly reported strategies involving fact-checking, cross-referencing, or using the LLM as a starting point. Students increasingly described using specific, detailed, and structured prompts to elicit more accurate or useful responses. 
In the pre survey, many students  described using LLMs to solve entire problems or complete assignments with minimal personal input and in the post survey, the use of these strategies decreased. There was a decline in “no verification” responses, and a rise in iterative questioning. Overall, `opening up the black box' and providing students with learning experiences of using LLM can empower students to develop responsible practices when working with LLMs.

%\section{Conclusion}
Our experience integrating LLMs into CS education suggests that students can develop a more nuanced, responsible, and technically grounded relationship with LLMs, and learn to partner thoughtfully, by  providing strategic inputs, scrutinizing outputs, and engaging in iterative refinement. Through `opening up the black box',
carefully designed assignments, reflective discussions about LLM use, and embedding verification loops,
CS educators can revise their courses to encourage students to seek their individual optimal approach to leveraging these tools in their CS journey.

%\begin{acks}
%Acknowledgments anonymized for the review.
%This material is based upon work supported by CETL ...
%\end{acks}

\bibliographystyle{ACM-Reference-Format}
\balance
\bibliography{sample-base}


\end{document}